\documentclass[sigconf]{acmart}
\usepackage{hyperref}
\usepackage[T1]{fontenc}
\usepackage[utf8]{inputenc}
\usepackage{microtype}
\usepackage{natbib}
\usepackage{url}
\usepackage{amsthm}
\usepackage[inline]{enumitem}
\usepackage{multirow}
\usepackage{color, colortbl}
\usepackage{xcolor}

\copyrightyear{2023}
\acmYear{2023}
\setcopyright{acmlicensed}\acmConference[CHIIR '23]{ACM SIGIR Conference on Human Information Interaction and Retrieval}{March 19--23, 2023}{Austin, TX, USA}
\acmBooktitle{ACM SIGIR Conference on Human Information Interaction and Retrieval (CHIIR '23), March 19--23, 2023, Austin, TX, USA}
\acmPrice{15.00}
\acmDOI{10.1145/3576840.3578309}
\acmISBN{979-8-4007-0035-4/23/03}

\begin{document}

\title{How Data Scientists Review the Scholarly Literature}

\author{Sheshera Mysore}
\email{smysore@cs.umass.edu}
\affiliation{%
 \institution{University of Massachusetts, Amherst}
 \country{USA}
}
\author{Mahmood Jasim}
\email{mjasim@cs.umass.edu}
\affiliation{%
 \institution{University of Massachusetts, Amherst}
 \country{USA}
}
\author{Haoru Song}
\email{hsong@umass.edu}
\affiliation{%
 \institution{University of Massachusetts, Amherst}
 \country{USA}
}
\author{Sarah Akbar}
\email{sakbar@umass.edu}
\affiliation{%
 \institution{University of Massachusetts, Amherst}
 \country{USA}
}
\author{Andre Kenneth Chase Randall}
\email{andrekenneth@umass.edu}
\affiliation{%
 \institution{University of Massachusetts, Amherst}
 \country{USA}
}
\author{Narges Mahyar}
\email{nmahyar@cs.umass.edu}
\affiliation{%
 \institution{University of Massachusetts, Amherst}
 \country{USA}
}

\renewcommand{\shortauthors}{Mysore, Jasim, Song, Akbar, Randall, and Mahyar}
\begin{abstract}
Keeping up with the research literature plays an important role in the workflow of scientists -- allowing them to understand a field, formulate the problems they focus on, and develop the solutions that they contribute, which in turn shape the nature of the discipline. In this paper, we examine the literature review practices of data scientists. Data science represents a field seeing an exponential rise in papers, and increasingly drawing on and being applied in numerous diverse disciplines. Recent efforts have seen the development of several tools intended to help data scientists cope with a deluge of research and coordinated efforts to develop AI tools intended to uncover the research frontier. Despite these trends indicative of the information overload faced by data scientists, no prior work has examined the specific practices and challenges faced by these scientists in an interdisciplinary field with evolving scholarly norms. In this paper, we close this gap through a set of semi-structured interviews and think-aloud protocols of industry and academic data scientists ($N = 20$). Our results while corroborating other knowledge workers' practices uncover several novel findings: individuals (1) are challenged in seeking and sensemaking of papers beyond their disciplinary bubbles, (2) struggle to understand papers in the face of missing details and mathematical content, (3) grapple with the deluge by leveraging the knowledge context in code, blogs, and talks, and (4) lean on their peers online and in-person. Furthermore, we outline future directions likely to help data scientists cope with the burgeoning research literature.
\end{abstract}


%
%

\maketitle

\section{Introduction}
\label{sec-problem-goals}
Literature reviews play an important role in the workflow of various scientists - with researchers spending upwards of 11 hours a week reading the research literature \cite{niu2010national}. They seek this literature with a variety of intents ranging from exploring the frontier of problems and solutions, keeping up their understanding, to quick look-up of facts, etc \cite{pain2016keep, Hoeber2019AcadSearch}. A challenge in this space is presented by an increasing volume of papers with studies estimating that the past few decades have seen a doubling of published research every 9 years \cite{landhuis2016infoverload, van2014global}. This deluge of information, as  \citet{Chu2021slowedcanonical} notes, presents a situation in which scientific progress as a whole sees a slowdown with individual scientists and peer-reviewers struggling to recognize and understand novel ideas. In this work, we examine the literature review practices of a currently emerging group of knowledge workers: data scientists.

Mirroring broader trends in the sciences, recent years have seen a surge in data science publications with a doubling every 2 years \cite{bollmann2020forgetting, neurips2019data, krenn2022predicting}, the development of several tools intended specifically to aid data scientists\footnote{\url{https://search.zeta-alpha.com/}, \url{https://papers.labml.ai/papers}, \url{https://alphasignal.ai/}}, and coordinated efforts to develop AI tools to identify the data science research frontier \cite{krenn2022predicting}. This hints at the information overload faced by this group. Further, data science as a discipline has had a significant impact in business \cite{forbes2013histds}, government \cite{persons2016data}, and science \cite{blei2017dsscience}, among others. Despite these trends, to the best of our knowledge, no prior work has examined the practices or challenges of data scientists as they review the research literature. Our goal is to address this gap.

Here, we view \textit{data scientists} as individuals engaged in data work, applied engineering, and research work and with training in computer science and statistics as well as disciplines of specific application domains such as economics and biology \cite{crisan2020passing}. 
We view \textit{literature reviews} as a knowledge-building process involving the gathering of research literature, development of relationships between gathered data, and the emergence of synthesized information --- spanning the tasks of \textit{information seeking}, \textit{sensemaking}, and \textit{composition} \cite{zhang2008citesense, sultanum2020litsense}. However, in this work, we omit examination of \textit{composition} given its narrower focus on publication and the difficulty of clearly demarcating these stages \cite[Sec 3]{vakkari2017salsys}.
\begin{figure*}[t]
  \centering
  \includegraphics[width=.6\linewidth]{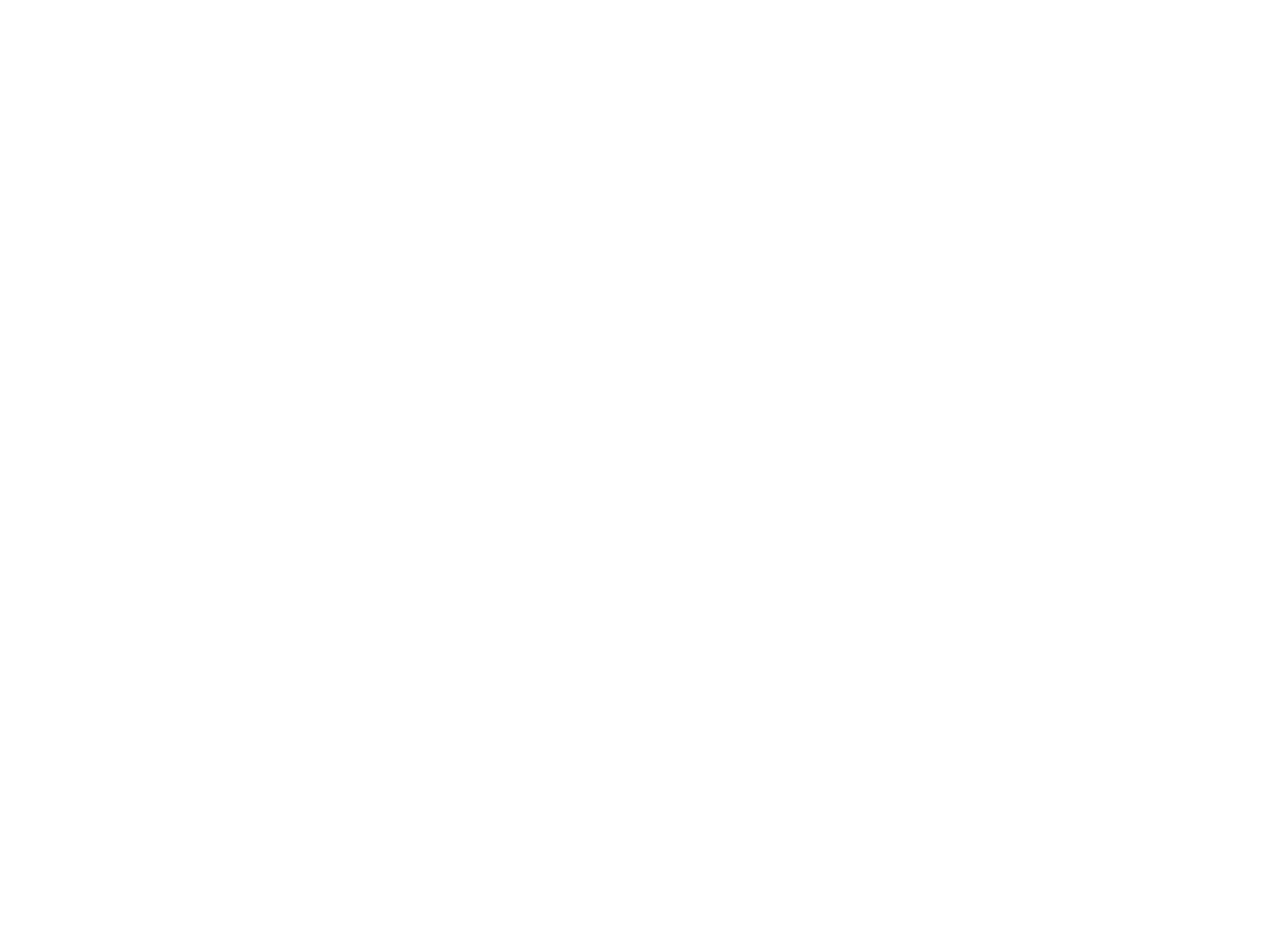}
  \caption{In this work, we examine the literature review practices of data scientists. We examine practices and challenges spanning information seeking and sensemaking: search and discovery, selection of literature, skimming and reading, and formation of a synthesized understanding of a collection of papers. \S \ref{sec-results} presents the results of our thematic analysis anchored to these stages. However, we omit the study of longer-term synthesis and composition.}
  \label{fig-teaser}
  \Description{Figure denoting the stages of the information search process - the process of query formulation and search, selection of sources, interaction with sources, and synthesis and composition of information.}
\end{figure*}

Specifically, we examine the practices and challenges in the iterative process of information seeking and sensemaking, where research literature is searched for, selected, read, and then synthesized into a mental model to achieve task-specific goals \cite{russel1993coststructure, white2009exploratory}.
In studying both information seeking and sensemaking as closely related activities, we draw on a line of work in \textit{search as learning} that views search as a learning process closely tied with sensemaking, and advocates support for users in both gathering information and use of the found results \cite{Kuhlthau1993, vakkari2017salsys, marchionini2018search, smith2019knowcontext}. Our study differs from the body of prior work which has examined practices of information seeking and sensemaking in \textit{isolation} -- focusing on search behaviors \cite{Athukorala2013compscilit}, use of academic social networks \cite{alhoori2019anatomy}, use of information management systems \cite{nosheen2018foundfound}, and note taking to capture ideas \cite{inie2022manageideas}. In cases when information seeking and sensemaking have been examined in conjunction it is in the context of specific literature-review systems \cite{sultanum2020litsense, matejka2021paperforager} or through the lens of abstract cognitive processes \cite{urgo2022pathway, anderson2001taxonomy} with limited focus on application grounded examination of literature review by expert knowledge workers.

Furthermore, in examining the literature review practices of data scientists our work also extends a recent body of work shedding light on the work practices of data scientists ranging from: teaching practices \cite{Kross2019dspractice}, code and data workflows \cite{Mash2021DSWorkFlow, Subramanian2020tractus, muller2019datawork}, to documentation practices \cite{Wang2021documentation}, and others. Our special focus on examining the practices of data scientists is also necessitated by prior work noting differences in literature review practices of scientists in different disciplines due to differences in research methods, norms of communication, and disciplinary values \cite{hilary2020disciplinary, Medlar2019absbody, alhoori2019anatomy, birhane2022values}.

To conduct this study we recruited 20 data scientists from industry and academic institutions and employed two different methods of gathering data for analysis: semi-structured interviews and think-aloud observations of open-ended search tasks conducted by participants. Our analysis used multiple rounds of coding and thematic analysis. In part, our study corroborates past findings, but also unearths several new and under-studied findings across information seeking and sensemaking: 1) Participants struggled in seeking literature, skimming, and establishing the credibility of literature outside their domains of expertise. 2) They struggled to understand papers in the face of missing details and mathematical content. 3) They faced an onslaught of seemingly similar papers and sought to understand them through their differences. 4) In coping with these problems they leveraged their peers and a range of online resources forming the \emph{knowledge context} of papers: discussion forums and social media, code, talks, and blogs. Given our findings, we highlight under-explored areas providing a rich canvas for future work and meaningful solutions for data scientists reviewing the literature.

\section{Related Work}
\label{sec-lit-review}
Here we overview four relevant lines of work. The first line of work overviews  \emph{information seeking} systems and practices for various knowledge workers. Alongside this, we overview the \emph{sensemaking} systems and practices for the scholarly literature. In examining the practices of knowledge workers we ground our findings of data scientists' practices in prior work. Similarly, our examination of existing systems combined with the challenges of data scientists (\S\ref{sec-results}) allows us to discuss meaningful future work (\S\ref{sec-discussion}). Following this work, we outline work on \emph{search as learning} which has examined the use of search systems in the context of learning tasks - much like literature reviews. Complementary to work on information seeking and sensemaking, this line of work also contributes important findings on the effects of tasks such as search, reading, and note-taking on learning-oriented goals. While these three lines of work have been explored in somewhat disjoint communities they may be seen as closely related, with each information seeking, sensemaking, and learning following one other \cite{marchionini2018search}. Finally, we overview the recent work examining \emph{data scientists} practices -- our work most directly extends this literature.

\subsection{Info-seeking \& Sensemaking - Practices}
\subsubsection{Information seeking.} 
A large body of work has studied the information-seeking practices of a range of different knowledge workers, ranging from medical researchers conducting systematic literature reviews \cite{knight2019biomedsys}, their use of search interfaces \cite{liu2022biomedinterfaces}, design students' use of search for ideation \cite{palani2021activesearch}, social science and education researchers practices \cite{Ince2018social}, astrophysicists practices \cite{sahu2013information}, and engineering professors perceptions of institutional library services \cite{debra2011libeng}. \citet{niu2010national} and \citet{alhoori2019anatomy} present extensive reviews of early work in this area. Given our focus on data scientists, we review studies that have examined information-seeking practices of populations in related disciplines \cite{Athukorala2013compscilit, alhoori2019anatomy, soufan2022searching, ishita2018parts, mccay2015exploratory}.

\citet{Athukorala2013compscilit} examine computer scientists for their goals in conducting literature searches and their tools. Their primary findings indicate that respondents used search most frequently to stay up to date on topics and find the exploration of unknown areas most challenging. They find that the most common form of navigating papers involves following citations to and from a known seed paper, often obtained via an initial keyword search. \citet{alhoori2019anatomy} examine STEM researchers' information-seeking behavior in conjunction with social media use and reference management tools. \citet{Hoeber2019AcadSearch} examine the differences in practices of STEM researchers across levels of seniority. Finally, recent work of \citet{soufan2022searching} presents a large-scale survey of STEM researchers intended to reveal the alignment between theoretical models of exploratory search and actual practices followed by researchers. 
A few studies have also examined behaviors grounded in specific systems: \citet{ishita2018parts} investigate which sections of a paper returned in search results researchers examined, while \citet{mccay2015exploratory} examine the difference between social scientists and computer scientists in using control features of digital libraries.

\subsubsection{Sensemaking} 
While a sizable body of work has examined information-seeking behaviors, a smaller body of work has examined practices in sensemaking tasks such as managing gathered literature, reading, and note-taking. \citet{nosheen2018foundfound} examines the challenges of personal information management in engineering researchers, finding fragmentation of data across different systems and difficulty determining the future value of information to be a challenge. Similarly, \citet{inie2022manageideas} explore researchers' tool use patterns in managing ideas, finding interoperability between tools to be important. Finally, \citet{morabito2021contextcap} studies researchers' use of tools to support the capture of context and metadata for documents to facilitate the effective reuse of found information. 

In contrast with this body of work which has examined information-seeking and sensemaking tasks in isolation, our work examines the two in conjunction through a series of interviews and think-aloud observations. This combined examination more faithfully examines the iterative processes of information seeking and sensemaking \cite{russel1993coststructure, white2009exploratory}, especially in the early stages of sensemaking. This allows our work to shed light on the practices and challenges at the intersection of these activities informing the development of systems for both activities.

\subsection{Info-seeking \& Sensemaking - Systems}
\label{sec-rel-work-systems}
\subsubsection{Information seeking} 
Prior work building information-seeking systems for the scientific literature has explored a range of strategies: discovery through seed papers and citation chaining \cite{li2021argoscholar, chau2011apolo, he2019paperpoles, ponsard2016paperquest}, through metadata such as authors or keyword \cite{dunaiski2017ConceptCloud, dork2012pivotpaths, portenoy2022bridger, kang2022cdqbe}, and systems exploring more specialized querying methods \cite{chan2018solvent, rahdari2021paperexplorer, choe2021papers101, kang2022augmenting, Michael2021DataHunter}. 

AgroScholar \cite{li2021argoscholar} and PaperQuest \cite{ponsard2016paperquest} facilitate visualization and interaction with the citation network starting from a seed set of papers - allowing incremental addition of papers to a core set of papers based on which others may be recommended. Work in Apolo \cite{chau2011apolo} and PaperPoles \cite{he2019paperpoles}, further supports user interactions such as grouping of papers and specification of preferences over keywords to influence the set of recommended papers. While this line of work has focused on citation-based discovery others leverage paper metadata more heavily. PivotPaths \cite{dork2012pivotpaths} presents a visualization system to allow researchers to explore paper collections through author and keyword metadata with a special emphasis on exploring the relationships between metadata elements and facilitating serendipitous discovery. Similarly, \citet{portenoy2022bridger} and \citet{kang2022cdqbe} enable the discovery of relevant authors and papers from disciplines different than that of the seed author or paper. Finally, a range of work has explored specialized querying methods going beyond papers or their metadata, more heavily leveraging the paper content. Work of \citet{chan2018solvent} and \citet{kang2022augmenting} allow searching for papers based on the ``challenges'', ``solutions'', or ``results'' they tackle and work in DataHunter \cite{farber2021datahunter} allows search of datasets based on research problem queries. 

\subsubsection{Sensemaking} We organize prior systems for sensemaking into aids for a \textit{collection} of documents \cite{heimerl2016citerivers, shahaf2012mapsofsci, matejka2021paperforager}, and systems to facilitate reading \cite{wang2016guidedlitr, zhang2021conceptscope, head2021augmenting, fok2022scim} and note-taking \cite{zyto2012socialdoc, subramonyam2020texsketch, kang2022threddy, han2022passages} of \textit{individual} documents . 

Since sensemaking often consists of iterative stages of information seeking and construction of mental models of gathered data \cite{pirolli2005sensemaking}, systems for sensemaking from collections of documents also contain components to aid information seeking - allowing users to examine and construct relationships while also allowing discovery of additional work \cite{chau2011apolo, dork2012pivotpaths}.  Other work has sought to help users determine salient time-aligned trends in the scientific literature. \citet{shahaf2012mapsofsci} construct ``metro maps'' from collections of scientific papers to depict the evolution of multiple lines of related work, similarly \citet{heimerl2016citerivers} leverage a streamgraph metaphor to allow users to examine patterns in citations, topics, and research communities over time. Finally, PaperForager  \cite{matejka2021paperforager} targets a different problem -- minimizing the context switches in document filtering, skimming, and reading. It represents paper collections entirely as persistently visible document images.

In augmenting reading experiences prior work has sought to help readers relate documents to other concepts and papers or have sought to enrich the reading experience of individual documents. \citet{zhang2021conceptscope} leverage bubble-tree maps to help visualize and explore hierarchies of concepts in a document. Similarly, \citet{wang2016guidedlitr} helps users better understand the related work sections of individual papers. On the other hand, reading aids for individual papers have examined aids for skimming through highlights over salient text \cite{fok2022scim}, while others have explored augmentations for reading equations and terms defined in a paper \cite{head2021augmenting}. Besides reading, work has also explored various forms of note-taking support to allow the synthesis and reorganization of gathered data. Work in Passages \cite{han2022passages} and Threddy \cite{kang2022threddy} help users to generate and organize text snippets and notes from reading papers that persist across applications and documents. While Passages emphasize persistent metadata and cross-application movement to retain the provenance of information, Threddy emphasizes the discovery of additional papers based on note collections. Finally, a line of work has also explored integrated systems for a literature review, spanning search and discovery, organization, synthesis, and composition \cite{sultanum2020litsense, zhang2008citesense}. Compared to work developing systems that study practices in the context of a specific system we conduct a more exploratory study examining scientists in their natural workflows aiming to contribute broader directions for future systems.



\subsection{Search as Learning}
The body of work on search as learning has studied several different aspects of how learning occurs in the context of the search process and provides several findings relevant to information-seeking and sensemaking. A bulk of the focus has been directed at studying learning outcomes while varying factors such as 1) user characteristics 2) search system characteristics, and 3) search behaviors \cite{vakkari2017salsys}. Here, learning has most generally been considered the formation of mental models of knowledge, their retention over time, and their application \cite{marchionini2018search}. 
This work also provides a perspective on search that goes beyond the finding and gathering of results. Instead, viewing search as a series of tasks in a constructive process - necessitating support for search as well as the use of search results \cite{Kuhlthau1993, vakkari2017salsys, marchionini2018search, smith2019knowcontext}. It is worth noting, however, that the bulk of work in SAL has focused on classroom students or drawn inferences with controlled experiments with crowd workers rather than knowledge workers such as data scientists. 

\subsubsection{User characteristics} Work on user characteristics has often examined users' domain knowledge. \citet{willoughby2009domaink} examine the outcome of domain knowledge alongside the use of search engines in an essay writing task. They find searching to only benefit essays where participants had greater domain knowledge, with participants noting domain knowledge to be beneficial for searching. \citet{roy2020salduring} examine learning with vocabulary assessments in the course of a search session as participants explore a topic, finding participants with greater prior topic knowledge to gain knowledge toward the end of their search session, while those without gaining more at the start of the session. While work examining user characteristics has often focused on domain knowledge \citet{vakkari2003termstactics} also find domain knowledge to influence searching only for users with knowledge of the search system.

\subsubsection{System characteristics } In examining system characteristics, prior work has explored search features \cite{sciascio2020ctrltransp, qiu2020memorableir, camara2021instscaffold} as well as features relating to reading and note-taking \cite{luanne2016textenvi, syed2020autoquestions, roy2021highlights}. Here, \citet{sciascio2020ctrltransp} find a transparent and controllable search system to deliver better learning outcomes than the widely used PubMed, and \citet{qiu2020memorableir} find web-search interfaces to lead to better knowledge gain compared to conversational search interfaces though the latter lead to improved knowledge retention. In studying reading experiences, \citet{luanne2016textenvi} find simpler text environments to lead to improved comprehension, \citet{roy2021highlights} find highlights in reading to improve topic coverage in essay writing tasks and note taking to lead to the inclusion of more facts, and \citet{syed2020autoquestions} finding automatically generated questions embedded in the text to improve learning outcomes.

\subsubsection{Search behaviors} Work examining search behaviors has contributed several findings. \citet{vakkari2003termstactics} find psychology students to use more specific queries as their vocabulary improved though their search strategies employed remained consistent. \citet{dosso2022queryform} find domain knowledge to not influence the number or length of queries across medicine and computer science students, though they find medicine students to use more domain-specific vocabulary in queries. \citet{vakkari2012effortdegrade} find increased effort in examining documents to be associated with improved essays even in the face of lower precision search results, finding students to compensate for bad search results -- a result also mirrored elsewhere \cite{smith2008useradaptation, liu2022biomedinterfaces}. Work of \citet{moraes2018instructor} finds search paired with instructor lectures to have improved learning outcomes than the lectures alone. Finally, \citet{urgo2022pathway} abstract away from specific interactions and characterizes the ``pathway'' toward learning objectives in terms of learning-oriented sub-goals for tasks varying in cognitive complexity. This also ties to a line of work examining task complexity and learning outcomes.

In our work, rather than focusing on specific learning outcomes and implementing interventions to facilitate learning, we contribute an exploratory needs-finding study for highlighting the practices and challenges throughout the process of search and the use of search results by data scientists in a learning-oriented task. In this respect, we draw on SAL by viewing search as a constructive process necessitating a focus on both information-seeking and sensemaking rather than the gathering of search results alone \cite{Kuhlthau1993, vakkari2017salsys}.

\subsection{Practices of Data Scientists}
\label{sec-ds-practice}
Besides the lines of work covered above, a sizable body of recent work has examined the broader practices of data scientists -- a currently emerging body of knowledge workers \cite{crisan2020passing}, we briefly discuss this work. Work of \citet{crisan2020passing} helps paint a picture of who data scientists are by conducting a review of prior work examining data scientists. A few papers have also examined the dataset-related information-seeking needs of data scientists. \citet{Kross2019dspractice} interview individuals engaged in training data scientists in industry and academia and identify that instructors find searching for pedagogically-relevant datasets to be one of their challenges. \citet{Koesten2017structdata} investigate data scientists' challenges in searching for structured data repositories. A range of work also examines data science workflows via codebooks \cite{Mash2021DSWorkFlow, Subramanian2020tractus}, their documentation practices \cite{Wang2021documentation}, and how they arrive at analysis decisions \cite{Kery2019foraging}. Other recent work examines the challenges and needs of data scientists in developing fair machine learning systems \cite{Holstein2019fairness}, and their adoption of explainable machine learning systems \cite{krishna2022disagreement}.

While this body of work has examined aspects of how data science is conducted, how data scientists seek and make sense of expanding research literature remains unknown -- as we have noted, the recent proliferation of tools aimed at data scientists and the expanding scholarly literature indicates that this is an important challenging activity. Therefore, in conducting our needs-finding exercise, our work serves to extend the understanding of the working practices of data scientists with a specific focus on information-seeking and sensemaking of the scholarly literature.

\section{Study Methods}
\label{sec-study-methods}
\begin{table*}[t]
\caption{Description of our participants. Research Areas were self-described in our semi-structured interview and our presentation respects participant requests for maintaining confidentiality about their research areas. Of 20 participants 13 were enrolled in Ph.D. programs in Universities across the USA and EU, and 7 participants held Data Scientist, ML Engineer, or Statistician designations at for-profit industry or non-profit organizations.}
\scalebox{0.85}{
\begin{tabular}{cllllllllllllllllll}
\toprule
Participant   &&&&&& Organization &&&&&& Designation &&&&&& Research Areas\\
\midrule
P1    &&&&&& Industry  &&&&&& Data Scientist &&&&&& Conversational Information Retrieval\\
P2    &&&&&& Non-profit  &&&&&& Statistician &&&&&& Statistics\\
P3    &&&&&& Industry   &&&&&& Data Scientist &&&&&& Computer Vision\\
P4    &&&&&& Industry   &&&&&& Data Scientist &&&&&& NLP for Legal Text\\
P7    &&&&&& Industry &&&&&&  Data Scientist &&&&&& Speech Processing\\
P5    &&&&&& University  &&&&&& Ph.D. Student &&&&&& ML for Public Health\\
P6    &&&&&& University  &&&&&& Ph.D. Student &&&&&& Video Processing\\
P8    &&&&&& University  &&&&&& Ph.D. Student &&&&&& Search as Learning\\
P9    &&&&&& University  &&&&&& Ph.D. Student &&&&&& Robustness in NLP\\
P10   &&&&&& University  &&&&&& Ph.D. Student &&&&&& Question Answering and Reasoning\\
P11   &&&&&& University  &&&&&& Ph.D. Student &&&&&& Robustness in NLP and Hate Speech\\
P12   &&&&&& University  &&&&&& Ph.D. Student &&&&&& NLP for Journalism\\
P13   &&&&&& University  &&&&&& Ph.D. Student &&&&&& Variational Inference\\
P14   &&&&&& Industry    &&&&&& ML Engineer &&&&&& NLP\\
P15   &&&&&& University  &&&&&& Ph.D. Student &&&&&& NLP for Scientific Documents\\
P16   &&&&&& University  &&&&&& Ph.D. Student &&&&&& Differential Privacy\\
P17   &&&&&& University  &&&&&& Ph.D. Student &&&&&& CS Theory\\
P18   &&&&&& Non-profit  &&&&&& Data Scientist &&&&&& Cryptography\\
P19   &&&&&& University  &&&&&& Ph.D. Student &&&&&& Reinforcement Learning and Causal Inference\\
P20   &&&&&& University  &&&&&& Ph.D. Student &&&&&& Data Management\\
\bottomrule
\end{tabular}}
\label{table-participant-descr}
\end{table*}

\subsection{Study Design} Our study primarily consisted of two components, a semi-structured interview followed by a think-aloud observation in the working environment of the participants' choice. While our semi-structured interviews probed participants' self-described practices and challenges, the think-aloud aimed to observe more tacit behaviors, workflows, and practices in a realistic task scenario. In our think-aloud participants conducted a literature review of their choice in response to one of three open-ended task prompts. The design and content of our study were developed in a pilot study with 15 participants run from February-May 2022. In our pilot study, we sought to probe information-seeking behaviors involved in literature review, however, in initial results we observed substantial sense-making activities closely tied to information-seeking. In the second iteration of our study with 20 new participants, run from July-August 2022, we updated our study to also probe broader sensemaking activities in literature reviews - our paper reports on this second iteration. Appendix \ref{sec-study-materials} includes our questions and think-aloud prompts.

\subsection{Study Participants} We invited participants for our study using social media posts on Twitter and LinkedIn, and email invitations sent to university mailing lists. In recruitment forms, a broad definition of "data scientist" was displayed and respondents were asked if they identified as data scientists. Further, participants were also asked to submit links/titles of 3 research papers they found useful or enjoyable to read in the past year. Of the respondents, 20 were selected as participants for our study (P1-P20) on a first-come-first-serve basis while ensuring that they identified as data scientists and had read papers in the past year. Participants were compensated for their time with a \$25 gift card and all study procedures and materials were approved by the university IRB. Appendix \ref{sec-study-materials} includes our recruitment form.

Of 20 participants 11 noted gender pronouns he/him, 9 noted she/hers, and 2 noted they/them, with some noting multiple pronouns. 14 participants were enrolled in or had completed Ph.D. degrees and 6 had completed master's degrees. 13 participants worked in universities and 7 worked in non-profit or for-profit industry labs. On average participants had published 4 research papers. Table \ref{table-participant-descr} describes participants' research areas and the type of organization where they conducted work.

\subsection{Study Procedure} Each study session was conducted by the authors and lasted about 1 hour with an equal split between the semi-structured interview and the think-aloud. The whole session was conducted over Zoom. We began each session by explaining the study procedure, obtaining participants' consent for participation, and having them fill out a short demographic survey. This was followed by a semi-structured interview. Here participants were asked about their research focus and their goals, practices, and challenges for conducting literature reviews. Participants were encouraged to think of literature reviews as broader than a single directed search and discuss all interactions with the scientific literature. This was followed by a think-aloud where participants conducted a literature review over Zoom screen-share. Participants were shown 3 broad task scenarios and selected one as a prompt for their think-aloud. One prompt asked them to recall and demonstrate a previous literature review, another asked them to enhance their knowledge on a known topic, and the final asked them to research the literature for a future project of their interest, an area they had lesser experience in. The latter two scenarios were drawn from the work of \citet{Hoeber2019AcadSearch}. Our analysis revealed that participants distributed uniformly between the 3 scenarios. Note that simple look-up searches were not used as task scenarios since we sought to understand more complex exploratory searches. If any papers were relevant enough to require in-depth reading participants were invited to move on and continue their search or terminate the study session. Audio and video of the study were recorded for analysis.

\subsection{Data Collection and Analysis} All our analyses used transcribed audio of the interview with the video examined for our think-aloud session. Automatic transcription of the audio was obtained from Zoom and was corrected substantially for errors. This was followed by 3 rounds of coding by 3 authors of this paper and a thematic analysis of the interview and think-aloud transcripts -- this process was conducted from Aug-October 2022. In our first round of coding 2 authors independently examined 5 transcripts and generated a set of codes using open and axial coding, then, these codes were consolidated into a unified set of codes. In consolidation, both authors examined each other's codes independently, then met to discuss any differences and consolidated them into a unified set of codes. This was followed by application of the unified codes to the 5 transcripts by both coders. We observed an agreement of $0.92$ in terms of nominal Krippendorff's alpha. This unified set contained 167 codes of which 51 noted names of tools used by participants or logistic aspects of the study. This was followed by discussion and resolution of disagreements in codes and application of the resultant codes to the remaining 15 transcripts while also adding any new codes to our initial codebook. This process added 59 new codes of which 32 noted names of tools or logistic aspects. The coded transcripts were then used for thematic analysis - which resulted in themes corresponding in part with the stages of the Information Search Process \cite[Sec 3]{vakkari2017salsys}, we present our themes next. Our codes and extended quotes per theme may be examined online.\footnote{Codes and extended quotes: \url{https://github.com/MSheshera/dslitreview-study}}

\section{Results}
\label{sec-results}
In presenting the results of our thematic analysis we broadly organize our themes by the Information Search Process as consisting of the formulation of an information need (\S\ref{sec-why-access}), query formulation and search (\S\ref{sec-how-access}), assessment of search results through skimming and reading (\S\ref{sec-how-read}, \S\ref{sec-how-select}), and synthesis of a collection of documents (\S\ref{sec-how-read}, \S\ref{sec-how-select}). Besides these, our final theme elaborates on how participants relied on social ties at various stages of the ISP (\S\ref{sec-how-social-ties}). Importantly, while these stages provide an organizing model to aid understanding, they are iterative processes with poorly defined boundaries. In presenting our results we present each theme followed by specific relevant prior work where appropriate.

\subsection{Why do data scientists access the scientific literature?}
\label{sec-why-access}
\subsubsection{Actively understanding disciplinary norms}
Participants most sought the literature actively as a means to understand disciplinary norms where they lacked this familiarity (18/20). Here, they sought to understand disciplinary vocabulary, form a mental model of the discipline, understand the contexts and community preferences of problems, solutions, and evaluation practices: 
\begin{quote}
    ``The first question that comes to mind for any researcher is what has been already done, are there similar problems which have already been tackled or related problems whose corresponding methods might be used in my problem.'' - P16
    
    ``When I'm starting work in a problem ... I'm not sufficiently familiar with to work to know what the typical approaches are, how is this evaluated, what kinds of approaches are falling out of favor versus becoming more accepted by the community.'' - P15
\end{quote}
Participants focus on understanding problems and solutions reflect prior understanding across scientific and creative problem-solving disciplines with individuals often thinking in terms of ``problem'' and ``solution'' \cite{heffernan2018identifying, palani2021activesearch}. A focus on evaluations and metrics also matches understanding of a large focus on quantitative performance in data science \cite{birhane2022values}. 

\subsubsection{Passively following a discipline} Besides actively seeking to understand a discipline, participants also noted accessing the literature more passively (10/20) -- keeping up with trends in the community or keeping updated on the work of peers: \emph{``Where the community is going, or what people that I have previously followed the works of are up to right now''} - P10, and remaining on the lookout for future projects: \emph{``Relevant reading for my own research what I had in mind ... at some point, I thought I would do future research in. [But] now I think that that's been on the back burner''} - P17. This motivation to keep up is also mirrored in related work \cite{Athukorala2013compscilit}.

\subsubsection{Brainstorming solutions} A majority of participants also leveraged existing scientific literature as a resource to aid brainstorming (18/20). Here participants leveraged the literature to find open problems (12/20): \emph{``Making sure I'm not redoing what has been done before, that's one major aspect -- figuring out like open questions and existing work that I could work on''} - P10. Others examined the space of solutions in prior work (13/20): \emph{``I am looking for existing methods which do what I am doing. What are other people doing to solve this? Is it even solved? Are there methods that already solve this problem?} - P16. While others established the novelty of their own ideas by seeking similar solutions (10/20): \emph{``After you figure out [the problem], it's like I have an idea for what you could do better, and then it's seeing if others have done something similar before''} - P5.

This process of seeking problems, developing solutions, and establishing the novelty of their ideas formed an iterative brainstorming loop. The use of search for creative brainstorming has been noted elsewhere \cite{palani2021activesearch}, and the desire for finding problems and solutions in the scientific literature has seen the development methods and resources intended to facilitate this \cite{chan2018solvent, kang2022augmenting, mysore2021csfcube} - our findings further support this effort. Focus on novelty and building on prior work also matches understanding their importance in data science research \cite{birhane2022values} and creative disciplines more broadly \cite{inie2022manageideas}.

\subsubsection{Seeking solutions for application} 
Besides seeking the literature for developing solutions, several participants also sought solutions to problems (17/20). Here, participants often sought solutions for direct application: \emph{``The other thing I will look this how other people have applied [this method], to see if other people have applied it to similar data or use cases to mine''} - P2, or as baseline systems to aid research: \emph{``What are the general approaches that people have taken to solve a given task... That helps when we are trying to publish a paper... So this forms what are the existing baseline methods to compare against for quantitative research''} - P6.

Seeking solutions as baselines for comparison or as solutions for application matches findings from citation analysis indicating that solutions either assume the role of vetted methods which are widely adopted or ones on the frontier of research knowledge and seeing current development, necessitating comparisons \cite[Sec 8]{jurgens2018citeframes}. Further, the acts of ``creating'' and ``applying'' also correspond to cognitive processes often used to precisely define learning objectives \cite{urgo2022pathway}.

\subsection{How do data scientists access the scientific literature?}
\label{sec-how-access}
\subsubsection{Data scientists seeking the literature} 
\label{sec-seeking-lit}
Expectedly, most participants sought the literature through keyword-based web searches (17/20), following citations to and from source articles (15/20) obtained via search or recommendations, or by examining papers of authors of relevant publications (13/20). This largely matches prior understanding \cite{Athukorala2013compscilit} and has seen the development of several tools intended to aid citation chaining behaviors \cite{chau2011apolo, ponsard2016paperquest, portenoy2022bridger}. Despite these methods being well established, participants often lamented the challenge of coming up with appropriate keywords to describe their information needs (13/20), sometimes requiring weeks to stumble into the right keywords. This was further exacerbated in unfamiliar disciplines:
\begin{quote}
    ``I had an idea in my head, but I was not sure how to map this into normative terms used by communities or even necessarily what communities are interested in this ... Ultimately it just took trial and error like finding some papers and coming back to it over several weeks, and eventually I kind of started to find things that actually matched.'' - P15
\end{quote}
To tackle this challenge in complex searches, prior work has explored query recommendation strategies \cite{palani2021conotate, medlar2021querysuggest, choe2021papers101}, with large-scale systems also implementing them \cite{namit2017gqsuggest}. More generally, challenges in formulating queries in corroborated in prior work noting literature searches to be exploratory with evolving, ill-defined, and open-ended searches -- where users often learn terms in the course of search \cite{soufan2022searching}. While many systems have been developed for exploratory search (see \S\ref{sec-rel-work-systems}), large-scale systems for literature search lack support for it \cite{nedumov2019exploratory}.

Finally, in the face of not knowing where to begin a search, participants also noted soliciting recommendations from expert peers (4/20): \emph{``If you don't know where to start and you just reach out to someone, especially in a company like [redacted] it's a lot easier, there are a lot of subject matter experts and they help us.''} - P7. As we note next support for leveraging social ties remains under-explored.

\subsubsection{The literature finding data scientists} 
\label{sec-lit-finding-ds}
Besides seeking the literature through search, participants also discovered scientific literature more passively (20/20). Their methods for obtaining automated recommendations primarily relied on following individuals on social media (13/20), subscribing to email alerts from arXiv or Google Scholar (7/20), following the work of specific authors (13/20), or newsletters intended to curate the literature (4/20). Here, however, participants noted being overwhelmed with the alerts which were supposed to inform them (5/20):
    \emph{``It would be nice to keep up with people's work ... Maybe at least you open up 10\% of this stuff instead of being like this is just junk mail.''} - P1.
Further, participants also noted tools for discovery often trapped them in a disciplinary bubble (4/20):
\emph{
    ``I'm probably heavily in my own bubble of papers ... if I'm working on hate speech, most of my recommendations will be very computer science based but maybe there's relevant stuff in social science that I'm probably never going to come across.''} - P11.

Besides automated methods for discovering literature, a majority of participants noted receiving recommendations from their peers (14/20) -- albeit less frequently than automated methods. This had several advantages - peers had a close understanding of participant interests, had vetted the paper, and offered deeper engagement:
\begin{quote}
    ``There are some of my peers from both industry and academia, who if they come across something interesting and they know [name] is working in these problems or she finds them interesting they just send it over. Sometimes it's, not even a paper it's just a blog post, which has links to papers.'' - P9
    
    ``I guess the benefit [of recommendations from peers is] its gone past one set of eyes. So there's some added incentive to read it, somebody said it was interesting and that the claims make sense.'' - P10
\end{quote}
Recent work of \citet{kang2022socialemail} finds even the addition of automatic social network-based explanations in scholarly email alerts to improve engagement and retention while allowing scientists to see themselves as part of the community.

\subsection{How do data scientists select papers?}
\label{sec-how-select}
\subsubsection{An onslaught of papers}
\label{sec-onslaught-of-papers}
Having obtained literature through search and discovery participants are now faced with filtering and organizing literature. Here they often reported being faced with a large volume of similar papers (16/20) and credited it to heavily crowded disciplines of data science. This made it hard to distinguish between many similar or incremental steps and seminal papers:
\begin{quote}
    ``If the field is very crowded - sometimes I find RL, and the problems I am focusing on to be crowded, then it becomes frustrating and you're always finding papers [that do the same thing].'' - P19 
    
    
    ``There are a lot of papers that are incremental updates so to find that one seminal paper that actually started it all is going to be painful and unless someone helps you out it's actually very hard.'' - P7
\end{quote}
To cope with this deluge participants turned to surveys or good reviews of the literature to find salient papers (7/20). Some choose to focus on a handful of papers, minimizing the overlap between papers they examined, and pairing the examination of a few papers with citation chaining to balance the breadth and depth of exploring results (9/20). Others also leveraged repeated references to specific concepts or papers as a sign of having found the papers worthy of examination (6/20): \emph{``Usually there will be a collection of papers cited in all of [the papers] and for me that is a good proxy of here is the core papers I should be looking at.''} - P16. In the presence of many similar variants, work in information foraging notes that users understand variants in terms of their differences or as time-aligned stories \cite{srinivasa2016simvariants}. While this is explored in the context of seeking code, we note a similar desire in the more challenging case of text documents, further elaborated on in \S\ref{sec-increments-progress}. 

\subsubsection{Establishing the credibility of papers.} 
\label{sec-establish-credibility}
An important challenge of coping with this deluge was establishing the credibility of papers (13/20). Here participants relied upon expected indicators: relying on known authors, affiliations, publication venues, and citation counts as markers of credibility. As we noted in \S\ref{sec-lit-finding-ds} some relied on the vetting provided in peer recommendations, others sought forum discussions: \emph{``One thing is that its hard to figure the credibility of a paper, so it's sort of trying to figure out the credibility based on discussions by online forums like Twitter, Reddit or Openreview. Even if this is highly reviewed what do other people who have worked in similar domains think about it''}- P14. 

Challenges with credibility however did not end with the selection of papers, participants also noted the content of papers sometimes betray their information scent (12/20). While this could in part be attributed to disciplinary writing norms \cite{huang2019holes}, participants also noted the challenge of papers with unclear or exaggerated contributions, re-branding of ideas, and papers not structured with information needs of a reader:
\begin{quote}
    ``People do a lot of re-branding, sometimes a lot of ideas are not very new but the motivation section is like poetry and when you read the details you feel [its] not what they are claiming they do. ... [or] exaggerating their contribution and not meeting the expectation in their experiments. So identifying those trends from papers is very important.'' - P19
\end{quote}
The use of publication metadata is well implemented in popular search engines, and its use for determining credibility has seen critical examination \cite{gomez2022leading, azzopardi2021searchbias}. However, methods and systems to facilitate the use of social discourse around papers to augment skimming and filtering have been under-explored. Further, while challenges like exaggerated contributions have been examined in science communication \cite{west2021scimisinfo} it remains understudied in literature searches by experts - we see evidence for this phenomenon here. These challenges arising from exaggerated/re-branded claims and needing to sift through many similar papers (\S\ref{sec-onslaught-of-papers}) may be a result of the crowded and highly incentivized disciplines of data science. While we present, to the best of our knowledge, the first report of these challenges they warrant deeper examination in data science and other rapidly expanding disciplines.

\subsubsection{Everyone skims papers.} Finally, in the selection of results all participants skimmed individual papers (20/20) often to make quick decisions of correctness or glean contributions of a paper. They often relied on knowing the discipline to know where to look for specific information and being challenged otherwise:
\begin{quote}
    ``I try to jump to wherever they say ``in this paper'' because I'm fairly familiar with the space I don't need to actually look at the introduction.'' - P7
    
    ``One challenge was that [this discipline] was very active in the 90s. ... They were published in different venues, and the approaches that they used were not familiar to me. That made it much harder to skim a paper and get the essence of its contribution.'' - P15
\end{quote}
Skimming is often interspersed with information seeking with frequent context changes between the two, however, few prior lines of work have examined close integration of information seeking and skimming \cite{matejka2021paperforager}, or tools to aid skimming of papers \cite{fok2022scim}.

\subsection{What challenges do data scientists face in reading papers?}
\label{sec-how-read}
\subsubsection{Understanding the hidden details}
\label{sec-hidden-details}
In reading and understanding papers, participants also noted the challenge born from missing details in a paper (13/20). Here they noted that papers were written to be accepted with little incentive for authors to include the details which made a solution effective:
\emph{
``Often the things that are in the paper or the things that make a good story, but the things that actually matter in the paper, for example, like how you set the hyperparameters are missing.''} - P5.  Some also noted a tension between including a lot of detail which would interfere with readers hoping to get a high-level idea of a paper: \emph{And then of course there's always a thing that I have the time to read a paper and you are getting into it and you face a lot of detail. Sometimes, thanks to the length and sometimes these certain details have to be omitted and, at times, those are the very details that would cause certain confusion to a reader.} - P9. In a bid to cope with missing details, participants noted the value of augmentations provided by code (7/20). 
\begin{quote}
    [I ask authors if] there is any publicly available code for what you're doing. Because many of these papers look well on paper but then its unclear how to implement them. Or its unclear which specific hyperparameter choices they made. - P16
\end{quote}
While platforms such as \url{paperswithcode.com}, pair papers and code, use of code to augment reading experiences for scientific papers presents a future venue for improvement. More broadly, the challenges arising from unclear documentation for models and data have led the data science community to pursue initiatives intended to document these artifacts \cite{mitchell2019modelcard, gebru2021datasheets}, sometimes incentivizing them at publication.\footnote{\url{https://naacl2022-reproducibility-track.github.io/}} However, even availability of code alongside papers remains at 25\% as of Sep 2022\footnote{\url{https://paperswithcode.com/trends}} with an understanding of incentives for documentation currently emerging \cite{chang2022mldoc}.

\subsubsection{Understanding the math on display}
While participants noted the absence of details the flipside was the inclusion of extensive math which also presented challenges (10/20). Here participants noted disciplinary subcultures which often rewarded papers with extensive math, equating hardness to understand with quality:
\begin{quote}
    In writing for niche audiences it requires having to show that [an idea] is important or useful and often that means that they will add equations or theorems [for an idea] that really are not as complicated ... if there's a lot of math or if it's hard to understand it must be impressive. - P5
\end{quote}
In coping with this, participants noted the importance of alternative sources such as code, blogs, and talks, or the inclusion of examples in papers to aid understanding (9/20). Also noting that talks and blogs presented better incentives to ensure understanding in audiences: \emph{Blogs help because sometimes the writing [in a paper] is not easy to understand compared to an informal way of writing [I: So maybe papers are more mathematical but blogs have the high-level ideas?] Exactly} - P6.

Aids for understanding math in reading papers have seen some precedent in recent work in the form of reading tools intended to better define mathematical symbols in papers \cite{head2022math, head2021augmenting}. However, this presents an open problem, with no work having explored augmentations from code, blogs, or talks to aid math understanding. It is also worth noting that challenges arising from missing details in papers and of understanding math, and the coping strategies of leveraging code and the knowledge context surrounding papers represent challenges that are likely specific to interdisciplinary and computational disciplines such as data science -- our work presents an initial report of this.

\subsubsection{Struggling to understand the increments of progress}
\label{sec-increments-progress}
In reading papers, participants often noted the importance of understanding the specific ``delta'' that papers offered in comparison to other work that was influential or known papers (9/20). This was important to forming an understanding of the discipline and contextualizing their contributions:
\begin{quote}
    ``Once you've read 10, 20, 30 papers it becomes a lot clearer what the core idea is and what the delta from the core is. ... For things you're not as well versed in, it's often a nightmare when you don't know the context in which the paper is being written. Understanding what exists in the literature and what doesn't is hard, then what the contribution is and why the contribution matters is hard.'' - P5
\end{quote}
To understand the increments participants noted the importance of gaining familiarity with the norms of a sub-discipline and the challenge in its absence: \emph{In grad school a professor synthesizes these things and says hey, this is the main theme of all these papers. When that information is there for you and you start reading the paper it tells you what to expect otherwise you're spending a lot of time and don't understand how different it is from previous papers''} - P7. 

However, it was a challenge to establish if a paper was poorly written or if the participants were missing necessary context: \emph{``The lack of clarity on their contribution - it might come from how they wrote that paper or it might come from my complete lack of understanding of what they might be doing''} - P19.

As we note in \S\ref{sec-onslaught-of-papers} in the presence of many variants of an item, prior work has noted users seeking to understand the differences between items \cite{srinivasa2016simvariants} - we see this here. Methods and systems to explore free text differences in documents to aid filtering, skimming and sense-making of collections remain under-explored. However, recent work in NLP has begun to examine this problem - \citet{luu2021explaining} explore methods for explaining relationships between papers, while \citet{rajagopal2022one} and \citet{kuznetsov2022revise} have studied textual edits in other application contexts.

\subsection{How do data scientists lean on social ties?}
\label{sec-how-social-ties}
Besides leveraging social ties for the discovery of papers in \S\ref{sec-lit-finding-ds}, participants also leaned on their peers in other ways.

\subsubsection{Collaboratively brainstorming and making sense of papers} 
All participants noted leveraging social ties in making sense of the literature (20/20). Here they noted the value of group discussions centered on papers to keep up with the literature, help brainstorm ideas, or spark new research directions (12/20):
\begin{quote}
    ``In independent research at the very least, I talk about my idea with someone else just to see if I'm in the right direction. On the other hand, we have bi-weekly brain-storming sessions in which we discuss at least one paper that is very relevant to our work and so we get a lot of inputs ... this might be useful, this might be a limitation, it might not work.'' - P7
\end{quote}
Others noted the value of discussions with collaborators to understand the details of specific important papers (11/20):\emph{``If I'm having one on one meetings then we dig into why people made certain decisions in their paper, and if we should be following the same''}  - P10. Finally, a few participants (5/20) noted the value of sharing notes and literature with collaborators to establish the provenance of ideas: \emph{I usually share these notes with collaborators to give them a sense of where I am getting this idea from or where this hypothesis is coming from} - P19 or correctness of information: \emph{It also helps with collaborators double checking my writing.} - P6.

Besides close peers, participants also sought weaker social ties and online discussions (14/20). Here, they noted many of the same benefits that interaction with peers provided - seeking recommendations from experts on forums: \emph{You will find a Quora or Yahoo answer which is dominated by exceptional mathematicians. Someone will have asked a similar question [to yours] and someone will have pointed them to some relevant work} - P17, establishing the credibility of papers as in \S\ref{sec-establish-credibility}, and engaged in discussion to understand the specifics of papers: \emph{[The ML Collective] discord channel has a lot of volunteers and contributors if you can figure out someone interested in the same kind of work ... you can be talking about how the algorithm goes from this step to another or what is this variable} - P14.

A body of work has examined collaborative information seeking \cite{morris2013collabsearch}, with some examining social reading \cite{zhang2021socialread, pearson2012coreading}, and sensemaking with collaborative annotations \cite{zyto2012socialann, shaikh2019web}.
However, examinations of collaborative reading and sensemaking in specific task contexts are understudied avenues that are likely to present specific challenges, as noted in cases of information re-use in software teams ~\cite{liu2021summknowledge} or collaboratively verifying the credibility of claims~\cite{lu2022debunkthis}.

\subsubsection{Leveraging authors}
While peers were useful in-person and online, participants also found value in interacting with authors (11/20). While some contacted them actively (5/20), others noted more passively seeking them in recorded talks and forums such as Twitter or Reddit (6/20). Direct communication ensured a fuller understanding of work, helped develop ties with other researchers, and sometimes turned into fruitful collaborations (2/20): \emph{A little while ago a new paper got released that was very similar to my work I emailed the authors. This is something I do in this type of situation to create a conversation and also connect with researchers who are doing similar work to me. I had a multi-day email chain with them where we discussed things and I wanted to confirm the differences [of my method] with them, which was extremely helpful} - P18. Others also found these useful to broach under-discussed aspects: \emph{I send them a message congratulating them about the work they've released and understand some of the secrets behind their work} - P4.

Here, prior work has explored recommending expert peers \cite{portenoy2022bridger} and ``people recommendation'' more broadly \cite{guy2016peoplerec}. However, important challenges remain in how incentive structures must be set up to facilitate these interactions e.g.\ \citet{breitinger2019ongoing} note a trade-off in the discovery of ongoing research and keeping it confidential.

While an active engagement was important for a deep understanding, passively interactions with authors through talks or forum posts made work easier to understand than their papers. Participants noted the value of visual communication and an incentive for authors to communicate their idea for understanding: \emph{Sometimes getting a very good intuition of what their motivation helps understanding. It might be not that clear in the paper but when they explain it to you in the video it's much more exciting. I think visual communication [is useful], their talk used a lot of good design and slide animations ... that cannot be communicated in a paper} - P19. However, some noted that talks were only recorded for ``famous people'': \emph{But [use of talks] isn't always applicable because it's only the more famous people's projects that get this much coverage. But I'm presuming I'm going into a brand new area and starting with the most famous people around} - P9.

Examination of authors' engagement through social media has only recently begun to be studied \cite{gero2021tweetorials, koivumaki2020social} and incorporation of this into aids for sensemaking remains under-explored. The usefulness of author interactions beyond their papers also indicates the value of alternative publication formats \cite{hohman2020communicating, iclr2022blogs}, once again, challenges remain in setting up incentives for publishers and authors \cite{team2021distill}.

\section{Discussion}
\label{sec-discussion}
In our Results, we examined the practices and challenges of data scientists reviewing the scientific literature, while anchoring our results along the formulation of an information need, query formulation and search, assessment of search results through skimming and reading, synthesis of a collection of documents, and leveraging social ties to accomplish a number of these tasks. Next, we note challenges that spanned across these themes and speculate future work likely to present solutions.

\subsection{Support cross-disciplinary access} 
Data science is seeing exponential growth in the number of papers -- a doubling every two years \cite{krenn2022predicting}. As fields develop several sub-disciplines emerge around specific problems and methods each with their own disciplinary vocabulary and norms \cite{jurgens2018citeframes}. Individual scientists are now tasked with conducting work in increasingly fragmented disciplines only some of which are familiar to them while knowing that significant innovation and progress is to be had from drawing across multiple disciplines \cite{andrey2015discovery}. The challenges stemming from fragmented knowledge are echoed in our findings at each examined stage of the information search process. Future work necessary to support information seeking and sensemaking of cross-disciplinary research can take several forms. 

To tackle the challenge of search in unknown disciplines aids such as query recommendation have been explored (\S\ref{sec-seeking-lit}), however querying strategies intended to allow richer specification of user context remain under-explored. These may take the form of verbose queries \cite{mysore2021csfcube, afzali2021pointrec}, conversational and interactive searches \cite{aliannejadi2021convsearch, handler2022clioquery} paired with mechanisms for cross-domain retrieval -- \citet{kang2022cdqbe} present a preliminary example. Further, the exploration of papers in unfamiliar disciplines is likely to benefit from the presentation of ``explanations'' to aid in understanding their credibility and relevance. While explanations have been extensively studied in recommender systems \cite{zhang2020explainable}, examination of explanations for cross-disciplinary exploration remains an open question. Aids may also be developed for skimming. Since skimming relies on knowing norms of writing in a discipline, these aids may take the form of adaptive document layouts \cite{jacobs2004adapt} or automatically generated ``FAQs'' for a paper \cite{august2022paper}, personalized to the discipline of a reader. Similarly, reading aids may take the form of paraphrasing text for different disciplinary audiences or presenting ``pre-requisite'' concepts or papers necessary to understand a given paper \cite{li2019should} -- thereby aiding readers to judge the quality of a paper independent from their own knowledge gaps. However, since scientists see better learning outcomes in tasks perceived as challenging the trade-offs involved in easing this process remains to be seen \cite{liu2022biomedinterfaces, vakkari2012effortdegrade}.

\subsection{Facilitate reliance on close peers} 
At several stages in reviewing the literature, participants relied on close peers --- teammates, lab members, collaborators, mentees, and mentors. From peers, they received recommendations, engaged in brainstorming, established the credibility of papers, and understood the details of papers. The collaborative practices of data scientists have been examined in the context of code and data work \cite{zhang2020dscollab} -- we extend this to include the creation and understanding of ideas. Careful understanding and support for these practices are likely to be fruitful -- especially with the move toward remote work \cite{yang2022effects}.

The dominant line of work in information-seeking has taken an egocentric perspective and exploration of methods and systems to leverage communities has been examined largely in early work in collaborative search \cite{morris2013collabsearch} and group recommendations \cite{jameson2007recommendation}. This line of work may fruitfully be re-explored in our present circumstance -- evidence for this is provided in recent work of \citet{piao2021friendrs} who find bringing ``friends-into-the-loop'' of recommender systems to result in more accurate and diverse recommendations. Re-incarnations of early work melding sensemaking, reading, and discovery in a collaborative feed-reader \cite{aizenbud2009coffeereader} also promise to leverage the trust scientists have in their peers for selection and consumption of research. A different line of work is suggested by \citet{morris2013collabsearch} where users preferred to interleave egocentric search with lightweight communication. Templates for this work are provided in aids to summarize group chats \cite{zhang2018groupchat} or collaborative conversational agents \cite{avula2018searchbot} -- as aids in brainstorming. Implementing these tools requires careful design, however -- \citet{brucks2022virtual} note virtual communication through video conferences to curb creative ideation. Similarly, the work of \citet{inie2022manageideas} notes users' preferences for their own scholarly tool chains, making interoperability of new tools important for uptake.

\subsection{Leverage the knowledge context of papers} 
While a close community of expert peers was important this was not always available to participants. As an expanding discipline with many new entrants, this is also likely in data science~\cite{janssen2022dstalent}. In this scenario, a variety of other resources were sought to augment papers -- forum and social media discussions, recorded talks and videos, blogs, and code. These resources provided starting points for exploration, helped establish the credibility of papers, and aided in understanding missing details and math in papers. Traditionally this information (e.g.\ entity cards, videos) referred to as the \emph{knowledge context} has seen use in web search engine result pages (SERP) and aids users in making information literate decisions. \citet{smith2019knowcontext} advocate for greater use of this knowledge context instead of their reduction and search engines ``getting out of the way'' of users. Here, we echo this push and observe its values for data scientists searching the literature. Furthermore, we also note its value in augmenting reading and skimming aids for found documents.

In the presentation of the knowledge context alongside academic search numerous questions remain. Prior work in the TREC Blog Track has examined blog retrieval with an emphasis on their subjective contents -- similar efforts are necessary for the retrieval of knowledge contexts for academic publications \cite{Ounis2008trecblog}. The presentation of this information is also likely to require further investigation, e.g.\ \citet{levi2018verticals} report that only some queries benefit from the presentation of a cluster of results from a Community Question Answering site. Further, questions remain about the fairness implications of augmenting academic search results with their knowledge context \cite{mcdonald2022search}. Besides use in information seeking, this knowledge context is likely to benefit skimming and reading aids. Recent work of \citet{rachat2022citeread} presents an early example and places discussions from follow-on work into the margins of a paper. There is room however for more complex augmentations -- This may take the form of using blogs, and videos for aiding math understanding, using code to infer missing details, or use of social media and blogs to aid skimming. This space remains under-explored with each type of information presenting challenges to retrieval, presentation, and the subsequent implications of use.

\section{Future Work and Limitations}
\label{sec-futurework}
We note three limitations to our study, each of which points toward future work. These stem from our study design, our choice of participants, and the nature of the challenges probed. First, given that our study spanned a single session we did not probe longer-running activities such as longer-term synthesis and composition, it is likely that examining this aspect is also likely to shed more light on the organizing and note-taking strategies of scientists. This will require longer-term observation and introspection from participants. Second, in the recruitment of participants we noted that all participants were based in the USA or EU with ample infrastructural access, it is likely that a more varied participant pool will result in different findings. Further larger-scale studies are also likely to be beneficial. Third, in several cases we noted challenges arising from incentive structures surrounding disseminating research, while reasonable in hindsight, these were under-examined in our study and warrant future work \cite{chang2022mldoc}. Finally, while unlikely to be a limitation, we also note that our study was conducted in the aftermath (February to October 2022) of the COVID-19 pandemic and its influence on work practices \cite{yang2022effects} -- this likely influenced our study and results.

\section{Conclusions}
\label{sec-conclusion}
An exponential rise in the number of scholarly publications, its expanding influence, a large number of new entrants, and its likely impact on data science drove us to examine information seeking and sensemaking practices of data scientists reviewing the literature. In our examination, we ran an exploratory interview study with 20 data scientists recruited from both industry and academic institutions. Here, we established their goals for accessing the literature, then we examined their practices and challenges in search and discovery, selection of search results, skimming and reading, and their reliance on peers in a number of these tasks. A number of our results corroborate those seen in prior work -- here, our work offers a synthesis of disparate prior work. Next, our work also uncovers specific results stemming from our focus on data scientists and our joint examination of information-seeking and sensemaking. In our findings, we highlighted challenges arising from fragmented scientific disciplines influencing all stages of the information search process -- we believe this represents a special challenge of data science given its inter-disciplinary nature. Besides, we highlighted the challenges of missing detail and mathematical content in reading papers -- likely, a feature of computational disciplines such as data science. Our joint examination of information seeking and sensemaking revealed the nature of information sought at the intersection of these two activities: a desire to establish the credibility of papers owing to exaggerated/re-branded claims or unfamiliarity with the discipline, and a desire to understand the incremental differences between many seemingly similar papers -- this likely represents a consequence of crowded and incentivized sub-disciplines of data-science. To cope with these challenges, we found our participants to leverage the knowledge context surrounding scientific papers in the form of code, blogs, talks, and forums and by leaning on their peers -- these practices were also examined in our work. Finally, examination of our participants' challenges and coping mechanisms lead us to carefully speculate on future work likely to present meaningful solutions to these challenges while also presenting meaningful scientific questions for future work in the IR, NLP, HCI, and CSCW communities.

\begin{acks}
We thank our study participants for their valuable insights and time. We also thank Alyx Burns of the HCI-VIS Lab for extensive guidance in his role as Teaching Assistant for the Advanced HCI class in the formative stages of this project. We acknowledge partial funding from the National Science Foundation under Grant Number IIS-1922090, the Chan Zuckerberg Initiative under the project Scientific Knowledge Base Construction, the National GEM Consortium, and the Intel Scholars Program.
\end{acks}

\bibliographystyle{ACM-Reference-Format}
\bibliography{sample-base}

\appendix
\section{Study Materials}
\label{sec-study-materials}
Here we present the recruitment form used for our study, the questions used in our semi-structured interviews, and the prompts used in our think-aloud. Additionally, we report participant demographics in Table \ref{table-participant-descr}

\subsection{Participant Recruitment}
\label{sec-recruitment}
The following contents were displayed in a web form to solicit information for participant recruitment:
\begin{enumerate}
    \item Your name. \emph{Response: Free text}
    \item Your email. This will help us to contact you for scheduling a study session. \emph{Response: Free text}
    \item Do you identify as a "data science" practitioner? Data science practitioners may be conceived as being very broad and inclusive of data work, applied engineering work, and research with individuals trained in computer science and statistics as well as disciplines of specific application domains such as economics and biology, among others. \emph{Response: Yes, No}
    \item Please list 3 research papers you have read in the past year and found enjoyable or otherwise useful. Please interpret "read" very broadly - this can include a full reading of the paper, skimming, or summaries read via blogs or tweets. \emph{Response: Free text}
\end{enumerate}

\subsection{Stage 1: Questions for Semi-structured Interview}
\label{sec-process-questions}
The following questions were asked in our semi-structured interview with follow-up questions (sub-points below) asked when appropriate and responses summarized to participants before the next question.
\begin{enumerate}
    \item Tell me a bit about your role as a data scientist. 
    \begin{itemize}
        \item For example: What kinds of problems do you work on? What is the nature of the work you conduct on a daily basis?
    \end{itemize}
    \item What are your goals in conducting a lit review?
    \item What kinds of supporting tools do you use in conducting your reviews? Eg. search engines, any bibliography managers, note-taking apps, etc.
    \item Are there points where you interact with others (virtually or in person) in the process of conducting reviews?
    \item Tell me about a time that you found conducting a literature review to be frustrating or tedious.
    \item Can you mention different ways in which you come across research literature? \emph{asked if time permits}
    \begin{itemize}
        \item What do you think the benefits and tradeoffs of those different methods are?
    \end{itemize}
    \item Before we conclude this stage, are there any additional thoughts about interactions with the literature that you would like to share?
\end{enumerate}

\subsection{Stage 2: Prompts for Think-Aloud with Task Scenario}
\label{sec-stage-1-think-aloud}
In this stage, participants conducted a literature review and shared their screens so they could be observed.

Participants were shown the following prompts and instructed to start their literature review process.
\begin{quote}
    ``Recall a literature review you conducted in the past. Imagine you were re-starting this process and show us how you went through the literature review.''
    
    ``Imagine you are interested in finding and documenting the latest work on a topic of your interest, show us how you go about this process.''
    
    ``Imagine you are planning future work on a problem you are interested in, conduct the literature review to help plan your future work.''
\end{quote}
\end{document}